\documentclass[trackchanges,twocolumn]{aastex7}
\usepackage{amsmath}
\usepackage{array}

\def\lowerbc{
\begin{figure}[ht!]
\includegraphics[width=\columnwidth]{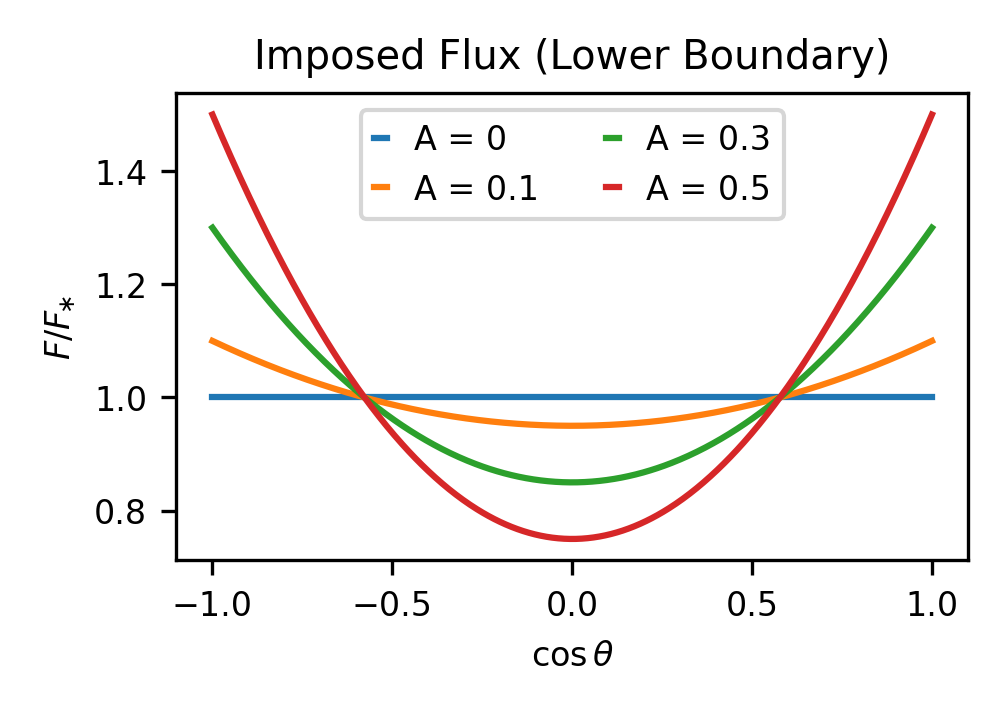}
\caption{Example profiles of normalized conductive heat flux $F/F_\star$ imposed at the lower boundary, where $F_\star=L_\star/4\pi r_i^2$.  Profiles are shown for a range of values of the forcing amplitude A \postref{as labeled}, corresponding to cases \postref{H1 to H4} in Table \ref{tab:input-output}.  Across our suite of models, the sign of A is chosen so that the latitudinal variation of flux at the lower boundary has the opposite sense of that which the system would otherwise establish when no variation is imposed. 
\label{fig:heatflux-bottom}}
\end{figure}
}

\def\fluxbalanceradius{
\begin{figure}[ht!]
\includegraphics[width=\columnwidth]{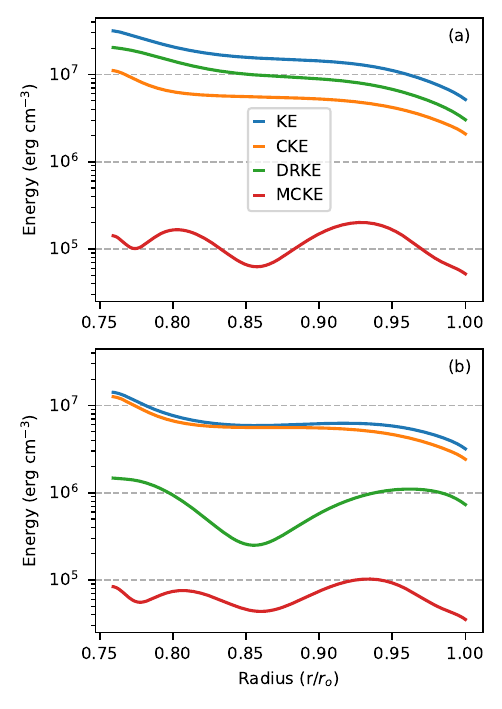}
\caption{Distribution of kinetic energy as a function of radius for a Ro$_\mathrm{c}$ = 1.75 model with no imposed flux (panel $a$) with an imposed flux amplitude $A=0.5$ (panel $b$), corresponding to cases \postref{H1 and H4} in Table \ref{tab:input-output}.  The different contributions to the kinetic energy as defined in Equations \ref{eq:cke} through \ref{eq:drke} are denoted by lines of different colors \postref{as labeled}.  As the thermal forcing increases to 0.5, the system shows a decrease in the energy associated with differential rotation (DRKE) and meridional circulation (MCKE), but the energy in the convective flows (CKE) remains nearly unchanged. 
\label{fig:E_v_rsadius_shell_Avgs_}}
\end{figure}
}

\def\fluxbalancetime{
\begin{figure}[t!]
\includegraphics[width=\columnwidth]{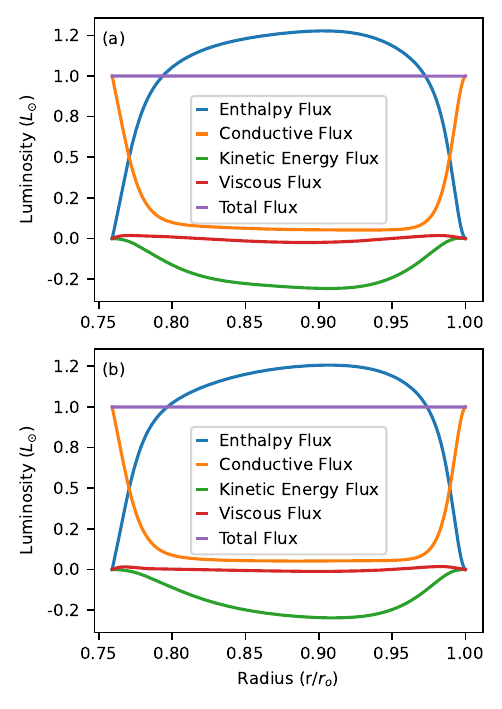}
\caption{Energy flux balance corresponding to the Ro$_\mathrm{c}$ = 1.75 models shown in Figure \ref{fig:E_v_rsadius_shell_Avgs_} with forcing amplitude $A=0$ (panel $a$) and $A=0.5$ (panel $b$).  Each contribution to the energy flux has been averaged in time and longitude and normalized by the solar flux $L_\odot/(4\pi r^2)$.  Labels indicate the different contributions as defined in Equations \ref{eq:enthalpy_flux} through \ref{eq:viscous_flux}. The spherically-averaged energy flux shows little change in the presence of strong thermal forcing.
\label{fig:shell_Avgs_lowflow}}
\end{figure}
}

\def\entropyDR{
\begin{figure*}[ht!]
\includegraphics[width=\textwidth, trim={0cm 0cm 0cm 0cm},clip]{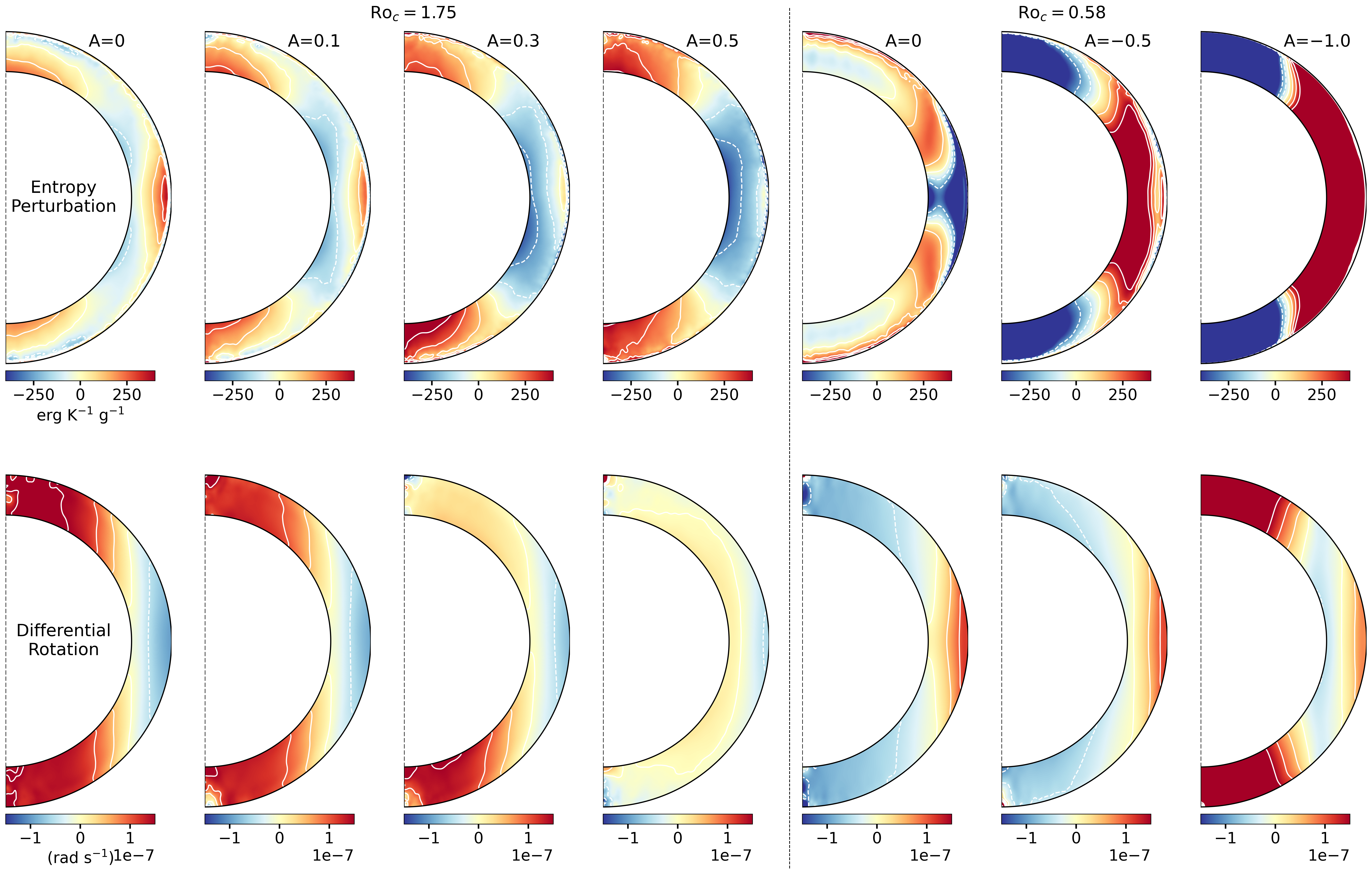}
\caption{Time and longitudinally-averaged profiles of specific entropy perturbation  (upper panels) and differential rotation (lower panels) for models subject to a range of forcing amplitude $A$. \postref{ A selection of Ro$_\mathrm{c}=1.75$} models are shown to the left of the vertical line, and Ro$_\mathrm{c}=0.58$ models at right.   As $A$ is increased, both \hrc ~and \lrc ~models exhibit a weakening of the differential rotation, and the latitudinal entropy variation increasingly reflects that imposed at the lower boundary. In the \lrc ~case, the system \postref{develops} an antisolar differential rotation when $A=-1.0$.
\label{fig:Entropy_DR}}
\end{figure*}
}

\def\thermalwindboth{
\begin{figure*}[ht!]
\includegraphics[width=\textwidth, trim={0cm 0cm 0cm 0cm},clip]{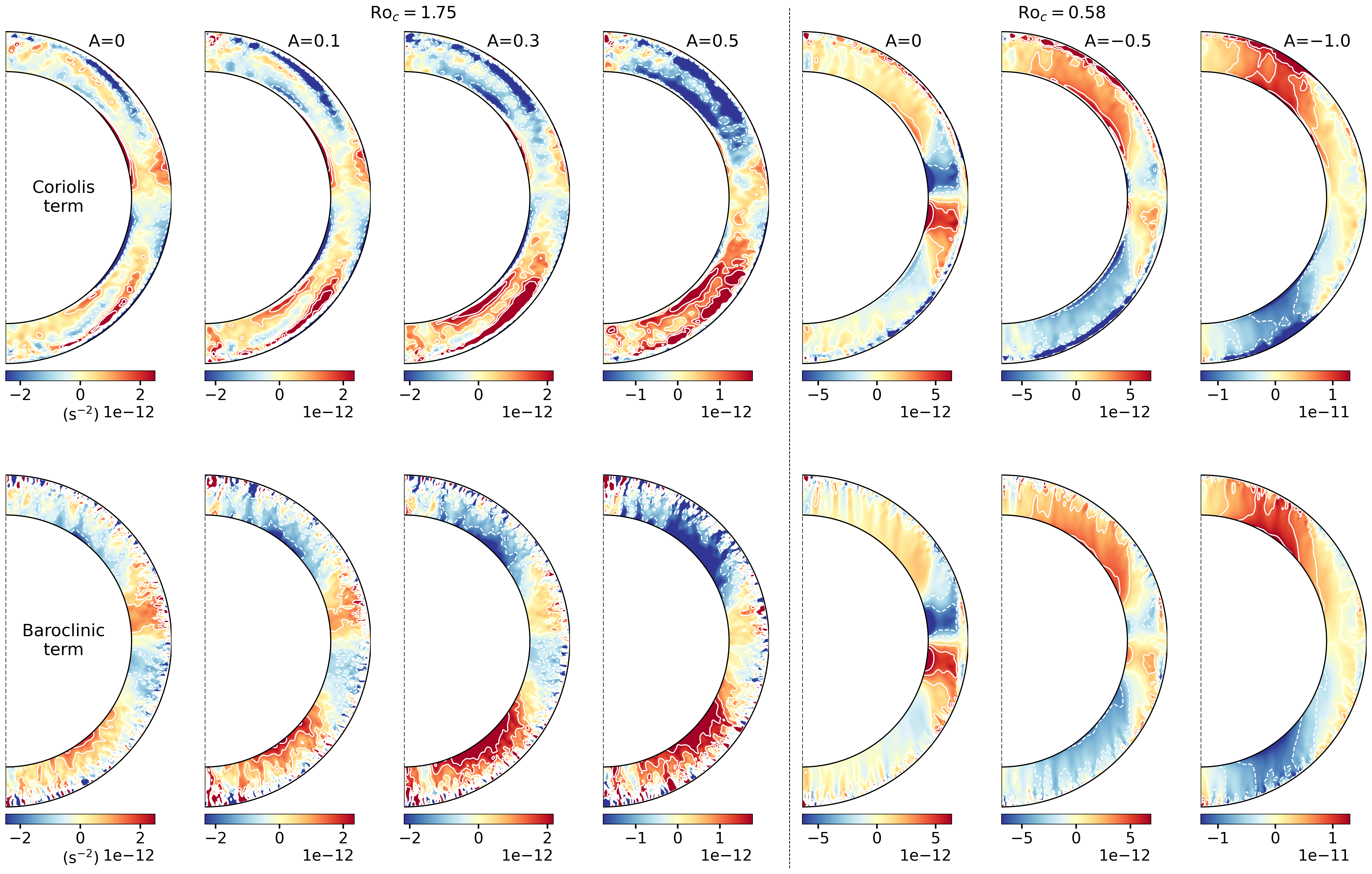}
\caption{Contributions to the thermal wind balance for those models illustrated in Figure \ref{fig:Entropy_DR}. Shown are \postref{(upper row)} the Coriolis contribution to the thermal wind balance, $2\Omega_0 r \sin\theta (\partial\Omega/\partial z)$   , and \postref{(lower row)} the baroclinic contribution, $g\,(r\,c_p)^{-1}(\partial s/\partial\theta)$.  In the absence of any imposed flux ($A=0$), there is a close correspondence between these two terms for the \lrc ~case.   This is also true of the \hrc ~system, though departures from thermal wind balance are apparent at high latitudes. In both systems, as $A$ is increased, the resulting configuration remains in thermal wind balance, but it \postref{is} modified relative to the $A=0$ case.   In particular, the contributions of the Coriolis and baroclinic terms increase at high latitudes and decrease at low latitudes.  
\label{fig:thermal_wind_high_Ro}}
\end{figure*}
}

\def\entropydrmc{
\begin{figure}[ht!]
\includegraphics[width=\columnwidth]{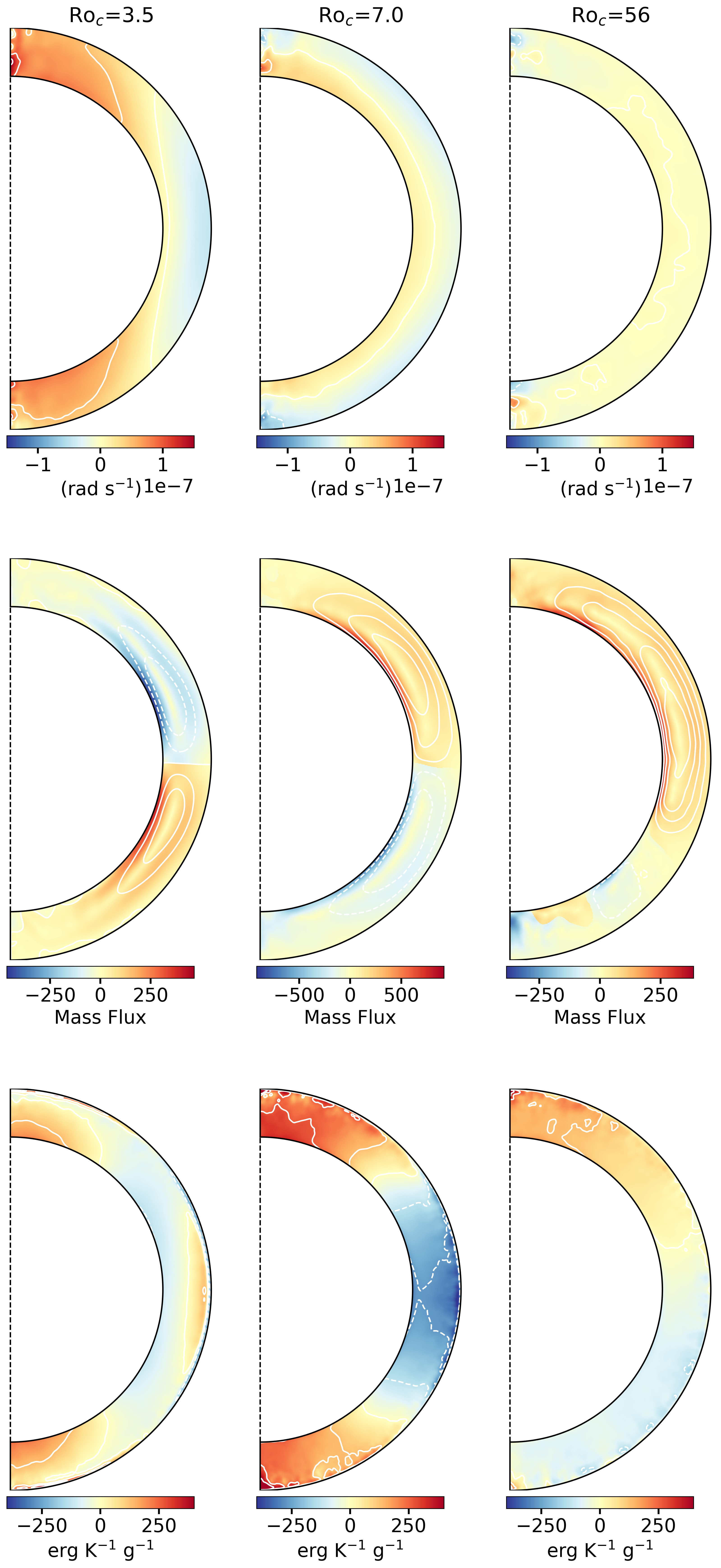}
\caption{Flow profiles at fixed forcing amplitude $A=0.3$ for varying Ro$_\mathrm{c}$. Time- and longitudinally-averaged (upper row) differential rotation (middle row) meridional mass flux and (lower row) specific entropy perturbation. Meridional circulation is colored red/blue for clockwise/counterclockwise flow. The equator-to-pole contrast in differential rotation declines with higher Ro$_\mathrm{c}$, eroding differential rotation. This decline is accompanied by an initial reversal in the sense of meridional circulation, which, along with the entropy perturbation, becomes non-axisymmetric. 
\label{fig:entropydrmc}}
\end{figure}
}

\def\thermalwindvaryro{
\begin{figure}[ht!]
\includegraphics[width=\columnwidth]{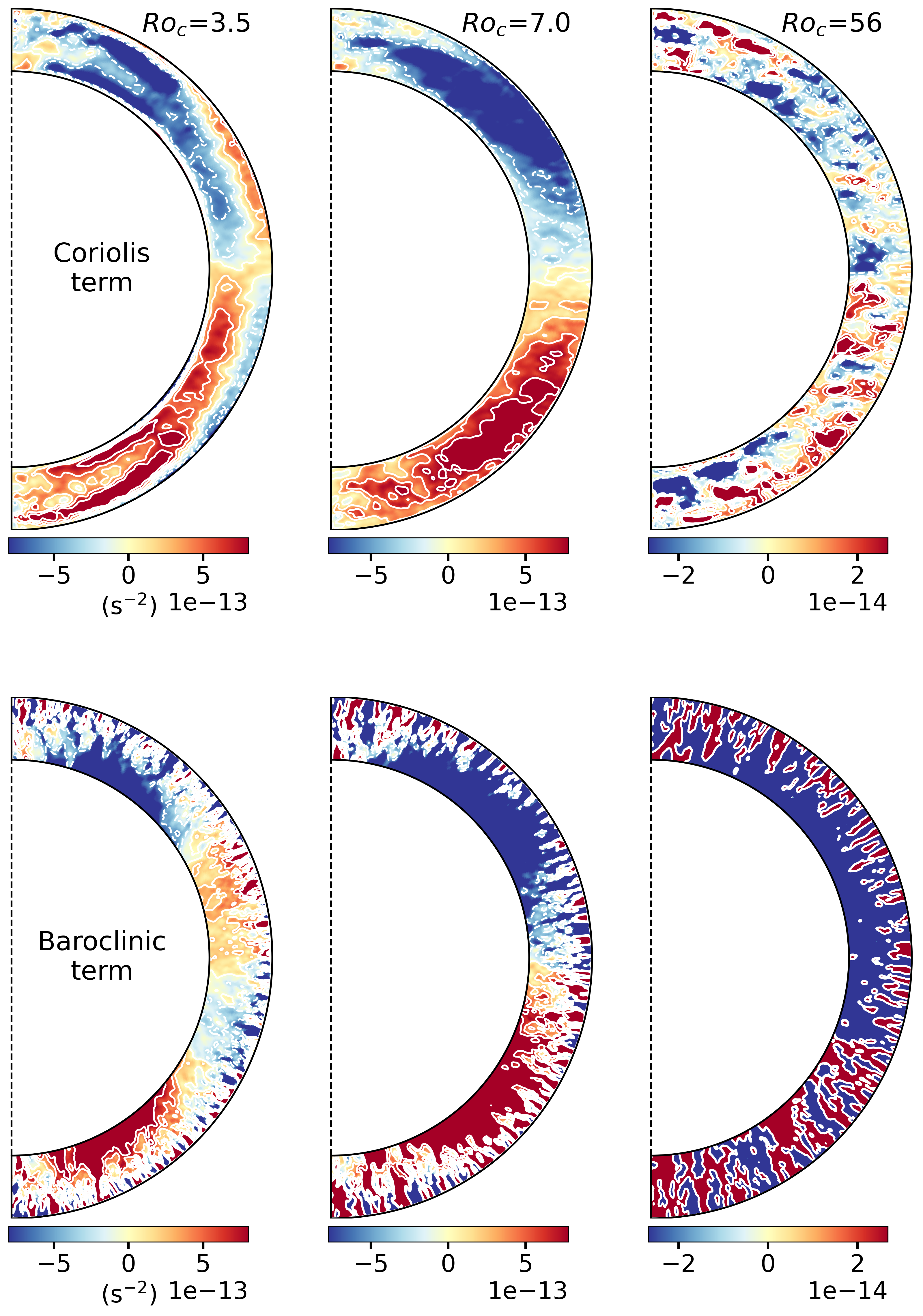}
\caption{Time- and longitudinally-averaged contributions to thermal wind balance for those models shown in Figure \ref{fig:entropydrmc}.  Shown are the contributions to thermal wind balance by \postref{(upper row)} the Coriolis term, and \postref{(lower row)} the baroclinic term. As Ro$\mathrm{c}$ increases,  the correspondence between these two terms progressively weakens, resulting in significant departures from thermal wind balance at high Ro$_\mathrm{c}$.
\label{fig:thermal_wind_vary_Ro}}
\end{figure}
}

\def\resultheatflux{
\begin{figure}[ht!]
\includegraphics[width=\columnwidth]{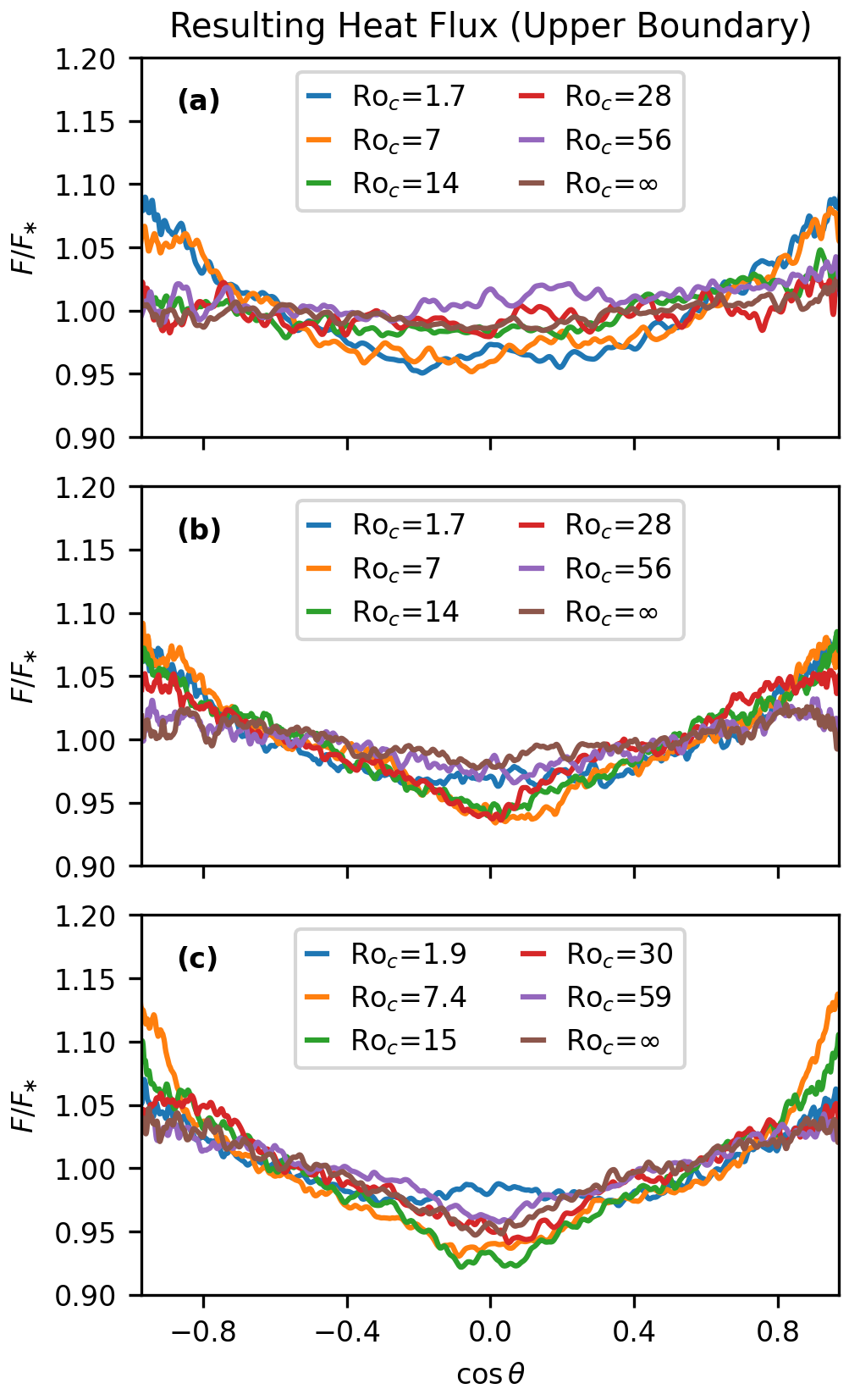}
\caption{Resulting conductive heat flux emergent at the outer boundary with fixed thermal forcing amplitude A = 0.3 \postref{and various values of Ro$_\mathrm{c}$}.  Shown are profiles for (a) and (b) thick- and thin-shell models, respectively, with Ra$_\mathrm{F}=1.355 \times 10^{6}$ and ($c$) thin-shell models with Ra$_\mathrm{F}= 3.845 \times 10^{5}$.   The imprint of the flux variation imposed at the base of the convection zone becomes weaker in all three series of models as Ro$_\mathrm{c}$ is increased.
\label{fig:heat_flux_top}}
\end{figure}
}

\def\tablepolytwo{
\begin{table}[h]
    \centering
    \renewcommand{\arraystretch}{1.2}
    \begin{tabular}{l l}
        \hline
        \multicolumn{2}{c}{Polytropic Background State Parameters} \\
        \hline
        $r_i$ & $5.00 \times 10^{10}$ cm  \\
        $r_o$ (thick shell) & $6.586 \times 10^{10}$ cm \\
        $r_o$ (thin shell) & $5.793 \times 10^{10}$  cm \\
        $M_i$ & $1.989 \times 10^{33} \text{ g}$ \\
        ${\rho}_i$ & $1.805 \times 10^{-1} \text{ g cm}^{-3}$ \\
        $c_p$ & $3.50 \times 10^8 \text{ erg K}^{-1} \text{ g}^{-1}$ \\
        $\gamma$ & 5/3 \\
        $n$ & 3/2 \\
        $N_{\rho}$ & 3 \\
        \hline
    \end{tabular}
        \caption{Summary of system parameters specifying the polytropic background state used here and described in detail in \citet{Jones2011} and \citet{Featherstone2016a}. Indicated are the values of inner radius $r_i$, the outer radius $r_o$ (which differs between thick- and thin-shell models), the mass interior to the lower boundary $M_i$, the density at the inner boundary $\rho_i$, the specific heat at constant pressure $c_p$, the adiabatic index $\gamma$, the polytropic index $n$ and the number of density scaleheights across the shell $N_\rho$.}\label{table:poly}
\end{table}
}

\def\inputoutput{
\begin{table*}[t!]
\centering
\begin{tabular}{cccccc|cccccc}   
\hline
& \multicolumn{5}{c|}{\textbf{Input parameters}}
& \multicolumn{6}{c}{\textbf{Output parameters}} \\
\cline{2-12}
\hline
Model & Ek & $\nu_{12}$ & $\Omega/\Omega_\odot$ & Ro$_\mathrm{c}$ & $A$  
& KE $(10^{6})$ & CKE $(10^{6})$ & DRKE $(10^{6})$ & MCKE & Ro & Re \\
\hline
\multicolumn{12}{c}{\textbf{Thick-shell, low-Ro$_\mathrm{c}$ models with Ra$_\mathrm{F}$=1.355$\times$10$^\mathbf{6}$ ; }$\mathbf{L_\star=L_\odot}$  } \\ 
\hline
L1 & 5.0 $\times 10^{-4}$ & 1.896 & 2.83 & 0.582 & 0.0 & 6.55 & 1.71 & 4.83  & 6.08 $\times 10^{3}$ & 0.04 & 77.19 \\
L2 & 5.0 $\times 10^{-4}$ & 1.896 & 2.83 & 0.582 & -0.1 & 6.06 & 1.74 & 4.32 & 6.11 $\times 10^{3}$ & 0.04 & 77.73 \\
L3 & 5.0 $\times 10^{-4}$ & 1.896 & 2.83 & 0.582 & -0.3 & 5.76 & 1.82 & 3.94 & 6.34 $\times 10^{3}$ & 0.04 & 79.37 \\
L4 & 5.0 $\times 10^{-4}$ & 1.896 & 2.83 & 0.582 & -0.5 & 5.78 & 1.95 & 3.82 & 6.91 $\times 10^{3}$ & 0.04 & 82.78 \\
L5 & 5.0 $\times 10^{-4}$ & 1.896 & 2.83 & 0.582 & -1.0 & 9.73 & 2.35 & 7.37 & 8.99 $\times 10^{3}$ & 0.04 & 89.35 \\
\hline
\multicolumn{12}{c}{\textbf{Thick-shell, high-Ro$_\mathrm{c}$ models with Ra$_\mathrm{F}$=1.355$\times$10$^\mathbf{6}$ ; }$\mathbf{L_\star=L_\odot}$  } \\ 
\hline
H1  & 1.5 $\times 10^{-3}$ & 1.896 & 0.94  & 1.746  & 0.0 & 14.70 & 5.14 & 9.43  & 1.26 $\times 10^{5}$ & 0.18 & 121.01 \\
H2  & 1.5 $\times 10^{-3}$ & 1.896 & 0.94  & 1.746  & 0.1 & 11.57 & 5.25 & 6.21  & 1.06 $\times 10^{5}$ & 0.18 & 120.84 \\
H3  & 1.5 $\times 10^{-3}$ & 1.896 & 0.94  & 1.746  & 0.3 & 11.19 & 5.27 & 5.83  & 0.93 $\times 10^{5}$ & 0.18 & 122.46 \\
H4  & 1.5 $\times 10^{-3}$ & 1.896 & 0.94  & 1.746  & 0.5 & 6.25  & 5.46 & 0.72  & 0.66 $\times 10^{5}$ & 0.19 & 126.06 \\
\hline

Tk1 & 3.0 $\times 10^{-3}$ & 1.896 & 0.47  & 3.492  & 0.3 & 8.04  & 5.79 & 2.08  & 1.84 $\times 10^{5}$ & 0.37 & 124.23 \\ 
Tk2 & 4.0 $\times 10^{-3}$ & 1.896 & 0.35 & 4.66 & 0.3 & 7.41 & 5.77 & 1.43 & 2.06 $\times 10^{5}$ & 0.50 & 125.76 \\
Tk3 & 6.0 $\times 10^{-3}$ & 1.896 & 0.24  & 6.985  & 0.3 & 7.26 & 6.47 & 0.47 & 3.11 $\times 10^{5}$ & 0.79 & 131.63 \\
TK4 & 8.0 $\times 10^{-3}$ & 1.896 & 0.18 & 9.31 & 0.3 &7.90 & 6.75 & 0.43 & 7.20 $\times 10^{5}$ & 1.08 & 134.75 \\
Tk5 & 1.2 $\times 10^{-2}$ & 1.896 & 0.12  & 13.960 & 0.3 & 7.82 & 7.11 & 0.25 & 4.66 $\times 10^{5}$ & 1.66 & 138.42 \\
Tk6 & 2.4 $\times 10^{-2}$ & 1.896 & 0.059 & 27.940 & 0.3 &7.90 & 7.35 & 0.25 & 3.05 $\times 10^{5}$ & 3.41 & 141.92 \\
Tk7 & 4.4 $\times 10^{-2}$ & 1.896 & 0.030 & 55.880 & 0.3 & 7.99 & 7.41 & 0.24 & 3.35 $\times 10^{5}$ & 6.78 & 141.19 \\
Tk8 & $\infty$ & 1.896     & 0.00 & $\infty$ & 0.3  & 8.08 & 7.50  & 0.25 & 3.34 $\times 10^{5}$ & $\infty$ & 142.47 \\

\hline
\multicolumn{12}{c}{\textbf{Thin-shell, high-Ro$_\mathrm{c}$ models with Ra$_\mathrm{F}$=1.355$\times$10$^\mathbf{6}$ ; }$\mathbf{L_\star=L_\odot}$  } \\ 
\hline
H5 & 1.5 $\times 10^{-3}$ & 0.946 & 1.89 & 1.746 & 0.0 & 53.39 & 6.07 & 47.21 & 1.16$\times 10^{5}$ & 0.19 & 127.05 \\
H6 & 1.5 $\times 10^{-3}$ & 0.946 & 1.89 & 1.746 & 0.1 & 46.80 & 5.55 & 41.14 & 1.03$\times 10^{5}$ & 0.18 & 121.57 \\
H7 & 1.5 $\times 10^{-3}$ & 0.946 & 1.89 & 1.746 & 0.3 & 49.62 & 6.18 & 43.36 & 0.84$\times 10^{5}$ & 0.19 & 124.69 \\
H8 & 1.5 $\times 10^{-3}$ & 0.946 & 1.89 & 1.746 & 0.5 & 68.76 & 6.58 & 62.09 & 0.86$\times 10^{5}$ & 0.20 & 130.35 \\
\hline
Tn1 & 3.0 $\times 10^{-3}$ & 0.946 & 0.942  & 3.492  & 0.3 & 9.50 & 6.68 & 2.72 & 0.95 $\times 10^{5}$ & 0.39 & 128.71 \\
Tn2 & 6.0 $\times 10^{-3}$ & 0.946 & 0.471 & 3.492  & 0.3 & 8.80 & 7.18 & 0.95 & 6.70 $\times 10^{5}$  & 0.80 & 133.58 \\
Tn3 & 1.2 $\times 10^{-2}$ & 0.946 & 0.236 & 13.970 & 0.3 & 9.44 & 7.72 & 0.49 & 12.30 $\times 10^{5}$ & 1.64 & 136.97 \\

Tn4 & 2.4 $\times 10^{-2}$ & 0.946 & 0.118 & 27.940 & 0.3  & 8.97 & 8.01 & 0.14 & 8.22$\times 10^{5}$ & 3.37 & 140.31 \\
Tn5 & 4.8 $\times 10^{-2}$ & 0.946 & 0.059 & 55.880 & 0.3  & 8.60 & 8.07 & 0.13 & 3.99$\times 10^{5}$ & 6.79 & 141.40 \\
Tn6 & $\infty$             & 0.946 & 0.00 & $\infty$ & 0.3  & 8.52 & 8.04 & 0.11 & 3.65 $\times 10^{5}$ & $\infty$ & 141.38 \\
\hline
\multicolumn{12}{c}{\textbf{Thin-shell, high-Ro$_\mathrm{c}$ models with Ra$_\mathrm{F}$=3.845$\times$10$^\mathbf{5}$ ; }$\mathbf{L_\star=0.284L_\odot}$  } \\ 
\hline
Tr1 & 1.5 $\times 10^{-3}$ & 0.946 & 1.89 & 0.930 & 0.3 & 3.56 & 1.41 & 2.15 & 0.07 $\times 10^{5}$ & 0.10 & 64.17 \\
Tr2 & 1.5 $\times 10^{-3}$ & 0.946 & 1.89  & 0.930  & 0.3 & 2.06 & 1.38 & 0.68 & 0.04 $ \times 10^{5}$ & 0.09 & 63.15 \\
Tr3 & 3.0 $\times 10^{-3}$ & 0.946 & 0.94  & 1.860  & 0.3 & 12.91 & 2.35 & 10.51 & 0.51 $\times 10^{5}$ & 0.24 & 78.42 \\
Tr4 & 3.0 $\times 10^{-3}$ & 0.946 & 0.94 & 1.860 & 0.3 & 8.04 & 2.33 & 5.68 & 0.32 $\times 10^{5}$ & 0.23 & 77.24 \\   
Tr5 & 6.0 $\times 10^{-3}$ & 0.946 & 0.47  & 3.720  & 0.3 & 2.90 & 2.64 & 0.23 & 0.30 $\times 10^{5}$ & 0.49 & 81.61 \\    
Tr6 & 1.2 $\times 10^{-2}$ & 0.946 & 0.23  & 7.440  & 0.3 & 3.81 & 2.90 & 0.44 & 4.78 $\times 10^{5}$ & 1.01 & 83.99 \\    
Tr7 & 2.4 $\times 10^{-2}$ & 0.946 & 0.11  & 14.881 & 0.3 &  3.91 & 3.12 & 0.12 & 6.70 $\times 10^{5}$ & 2.09 & 86.92 \\   
Tr8 & 4.8 $\times 10^{-2}$ & 0.946 & 0.059 & 29.763 & 0.3 & 3.49 & 3.22 & 0.06 & 2.11 $\times 10^{5}$   & 4.29 & 89.41 \\  
Tr9 & 9.6 $\times 10^{-2}$ & 0.946 & 0.029 & 59.526 & 0.3 & 3.48 & 3.24 & 0.05  & 1.87 $\times 10^{5}$  & 8.61 & 89.74 \\  
Tr10 & $\infty$            & 0.946 & 0.000 & $\infty$  & 0.3 & 3.46 & 3.24 & 0.05 & 1.69 $\times 10^{5}$ & $\infty$ & 90.06 \\
\hline
\end{tabular}
\caption{Input and output parameters for each model considered in this study.  The kinematic viscosity, $\nu_{12}$, is reported in units of 10$^{12}$ cm$^{2}$ s$^{-1}$. The rotation rate of each model, $\Omega$, is reported relative to the solar rate of $\Omega_\odot = 2.66\times 10^{-6}$ rad s$^{-1}$.   The left-hand columns also indicate the Ekman number, Ek, the convective Rossby number, Ro$_\mathrm{c}$, and the forcing amplitude, $A$, adopted for each model.  The right-hand columns indicate the time-averaged resulting values of kinetic energy density associated with the total flow, KE, the convective flow, CKE, the differential rotation, DRKE, and the meridional circulation, MCKE, quoted in units of erg cm$^{-3}$.  In addition, the resulting system-scale Rossby number, Ro, and Reynolds number, Re, are quoted for each model.  The luminosity of each model, $L_\star$, is reported relative to the solar value of $L_\odot = 3.846 \times 10^{33}$ erg s$^{-1}$.  
}
\label{tab:input-output}
\end{table*}
}

\newcommand{\bvec}[1]{\boldsymbol{#1}}
\newcommand{\mbvec}[1]{\mathcal{\boldsymbol{#1}}}
\newcommand{\hrc}{high-Ro$_\mathrm{c}$}
\newcommand{\lrc}{low-Ro$_\mathrm{c}$}
\newcommand{\postref}[1]{{#1}}

\graphicspath{{./}{figures/}}

\begin{document}

\title{The Response of Rotating Stellar Convection to Latitudinally-Varying Heat Flux}

\author[orcid=0009-0001-3457-0840,sname='North America']{Kinfe Teweldebirhan}
\affiliation{Southwest Research Institute, Department of Solar and Heliospheric Physics, Boulder, CO 80302, USA}
\affiliation{Aksum University, Department of Physics, Axum, Ethiopia}
\email[show]{kinfe.gebreegzabihar@swri.org}  

\author[gname=Savannah,sname='North America']{Rituparna Curt}
\affiliation{W.W. Hansen Experimental Physics Laboratory, Stanford University, Stanford, CA 94305-4085, USA}
\email{ritucurt@stanford.edu}

\author[orcid=0000-0002-2256-5884,gname=Bosque, sname='North America']{Nicholas A. Featherstone} 
\affiliation{Southwest Research Institute, Department of Solar and Heliospheric Physics, Boulder, CO 80302, USA}
\email{nicholas.featherstone@swri.org}

\begin{abstract}
We investigate how rotating convection responds to the imposition of a latitudinally-varying heat flux \postref{at the base of the convective layer}.  This study is motivated by the solar near-surface shear layer, whose flows are thought to transition from a buoyancy-dominated regime near the photosphere to a rotation-dominated regime at depth.  Here, we conduct a suite of spherical 3-D, nonlinear simulations of rotating convection that operate in either the buoyancy-dominated (high-Rossby-number, high-Ro) or rotation-dominated (low-Rossby-number, low-Ro) regime.  \postref{At the base of each model convection zone, we impose a heat flux whose latitudinal variation is opposite to the variation that the system would ordinarily develop.}  In both \postref{the low- and high-Ro} regimes, a strong thermal wind balance is sustained in the absence of forcing.  With a larger flux variation, this balance becomes stronger at high latitudes and weaker at low latitudes.  \postref{The resulting differential rotation weakens in response and, at sufficiently high forcing, its latitudinal variation reverses for both low- and high-Ro systems.}  At fixed forcing, there exists a Rossby number above which the convective flows efficiently mix heat laterally, and the imposed flux variation does not imprint to the surface.  At sufficiently high-Ro, thermal wind balance is no longer satisfied.  We discuss these results within the context of the Sun's near-surface region, which possesses a weakened differential rotation when compared to the deep convection, along with little-to-no variation of photospheric emissivity in latitude.  
\end{abstract}


\section{Introduction} 
Helioseismology has revealed the detailed angular velocity distribution within the solar interior. These observations show that the \postref{Sun's convection zone} exhibits differential rotation, with a rapidly-rotating equator and slowly-rotating poles.  \postref{The convection zone} also possesses layers of strong radial shear at its base (the tachocline) and near the photosphere \citep[the near-surface shear layer \postref{(NSSL)}; e.g., ][]{Thompson2003, Howe2005, Howe2009, Barekat2014, Barekat2016, Komm22, Antia2022, Rabello2024, Sen2025}.   The strong radial shear observed in these regions, and its potential role in the dynamo process, has made both the tachocline and the NSSL a central focus of numerous observational and theoretical studies \citep[e.g., ][]{Charbonneau2020}.  In the present work, we focus on the NSSL and explore how its rapid flows might impact the mixing of heat in the Sun's near-photospheric layers.


Several efforts have been undertaken to study the NSSL numerically.  These range from early, thin-shell simulations to later, deep-shell simulations that incorporated significant background density stratification and, recently, magnetism \citep{Gilman1979, Gastine2013, Guerrero2013, Hotta2015, Matilsky2019, Hotta2025}. One common thread from these investigations is that NSSL-like behavior emerges in part through the interaction of rapid, near-surface flows, which are weakly influenced by the Coriolis force, with slow, deep convective flows that sense it more strongly.

The relative importance of the Coriolis force is often characterized through a Rossby Number (Ro) that expresses the ratio of \postref{the rotation period to the convective timescale}.  A number of numerical studies have now demonstrated the usefulness of Ro \postref{ for determining the characteristics of the differential rotation} that is established \citep{Gastine2013, Gastine2014, Guerrero2013, Featherstone2015, Camisassa2022}.  When Ro is low, and the Coriolis force is significant, convection produces solar-like differential rotation characterized by a fast equator, slow poles, and a tendency toward slightly warmer polar latitudes.  In contrast, at high Ro, convective systems transition to a so-called ``antisolar'' regime with slow equatorial rotation, fast poles, and warmer equatorial regions.

These results, combined with the rapidly-rotating equator observed in the Sun, would seem to suggest that the bulk of the convection zone is operating in a low-Ro regime.  And yet, helioseismic measurements of the deep structure and amplitude of solar convection remain inconclusive, with many studies yielding conflicting results \citep[e.g.,][]{Hanasoge2012,Greer2015,Proxauf2021,Birch2024}.  \postref{When flows are observed in the photosphere and the near-photospheric layers of the convection zone}, however, the picture becomes much clearer.  The dominant scale of motion in this region of the Sun is supergranulation \citep{Hart1956,Leighton1962,Rincon2018}.  Supergranulation is characterized by motions with spatial scales of roughly 30,000 km, flows speeds of roughly 400 m s$^{-1}$ and lifetimes of less than one day \citep[e.g., ][]{Rast2003}.  These numbers suggest that the convective timescale of supergranulation is considerably shorter than the Sun's rotational period, which raises an interesting point. \postref{It is often asked why the Sun's differential rotation weakens in the NSSL}. \textit{We might equally ask why it does not reverse its sense of equator-to-pole variation all together.}

Another apparent inconsistency arises when considering observations of the Sun's photospheric emissivity. Multiple measurements demonstrate that pole-to-equator temperature variations are limited to only a few Kelvin \citep{Falchiani1974, Kuhn1998, Kuhn1998b, Rast2008}. In contrast, global models of low-Ro rotating convection typically observe a flux variation of order 10$\%$ between equator and pole, which would translate to a temperature contrast of $\mathcal{O}(100)$ K \citep[e.g.,][]{Featherstone2015}.  Alternative theoretical arguments based on geostrophic balance (i.e., a balance between pressure and Coriolis forces) arrive at a smaller, but still too large, number of $\mathcal{O}(10)$ K \citep{Matilsky2020}. These discrepancies raise a second question that motivates this work. \textit{What role, if any, might high-Ro flows in the NSSL play in the latitudinal mixing of heat?}



New progress on these questions concerning the nature of differential rotation and the temperature contrast \postref{observed in the photosphere} has now been made through a theoretical model described in  \citet{Choudhuri2021, Jha2021}.  That model assumes that the NSSL is in thermal wind balance, meaning that the dominant force balance struck on large scales occurs between buoyancy, Coriolis and Pressure forces \citep[e.g.,][]{Rempel2005,Miesch2006}.  This assumption, combined with the assumption that surface latitudinal temperature variations persist in depth across the NSSL, with no lateral mixing, leads to a predicted near-surface differential rotation that is well-matched by observations.


And yet, the success of this model introduces a puzzle of its own. The assumption of geostrophy underpinning thermal wind balance is typically associated with systems in which the Coriolis force is significant.  \postref{It is expected for low-Ro systems,} but its relevance to high-Ro systems such as the NSSL is much less clear. \textit{We might then ask under what range of conditions a high-Ro system can sustain thermal wind balance.}  

In this work, we explore these and related questions concerning the mixing of angular momentum and heat by simulating the response of rotating convection to an imposed, latitudinally-varying heat flux.  Through these simulations, we explore how the lateral mixing of heat, and the resulting thermal wind balance, respond to the Rossby number of the convection and the strength of the imposed flux variation. 

\section{Numerical Experiment}
\label{sec:methology}
We now describe a set of spherical 3-D numerical simulations of rotating convection designed to resemble a high-Ro layer of convection overlying a low-Ro interior, and vice versa.  This setup is motivated by the fact that the solar convection zone appears to be operating in both the high-Ro and the low-Ro regimes.  Specifically, the rapidly-rotating equator suggests that the bulk of the convection zone operates in a low-Ro regime and should preferentially transport heat more efficiently at high latitudes.  \postref{However,} the observed photospheric emissivity has essentially no latitudinal variation.  At the same time, the near-photospheric flows are observably high-Ro in nature and would tend to drive a slowly-rotating equator while transporting heat more efficiently at the equator.

The models presented here span both low- and high-Ro regimes.  For each model considered, we impose a latitudinally-variable heat flux at the lower boundary that opposes that which the system would naturally establish. In low-Ro systems, we impose a flux that is stronger at the equator and weaker at the poles, naturally driving an equatorial region that is warm relative to the poles.  This opposite is true for high-Ro systems, where we impose a flux that is stronger at high latitudes and weaker at low latitudes.  

We consider ``high-Ro'' to mean a \postref{convective Rossby number}, Ro$_\mathrm{c}$, greater than unity, while ``low-Ro'' indicates a value of Ro$_\mathrm{c}$ that is less than unity.  A value of one for this system control parameter, whose precise definition we defer until the discussion \postref{in} \S \ref{sec:nondim}, has been shown to clearly delineate the transition between solar and antisolar differential rotation \citep{Gastine2014,Camisassa2022}. Across our suite of simulations, we examine how heat is mixed in latitude within the fluid shell.  We also examine how the resulting differential rotation and thermal-wind balance respond to the presence and strength of this thermal forcing.

\subsection{Anelastic Formulation of the Fluid Equations}
All simulations presented in this study have been modeled under the anelastic approximation. This treatment is well-suited for deep stellar interiors where both the fluid Mach number and deviations from an adiabatic background atmosphere are expected to be small \citep[e.g.,][]{Batchelor1953, Gough1969, Gilman1981}.  The anelastic formulation is particularly well-suited for the study of stellar interiors as it admits strong depth-variation in the background parameters, such as density, while also filtering out sound waves that otherwise severely limit the computational time step.

Within this framework, all thermodynamic quantities are defined with respect to a background state that satisfies the ideal gas law:
\begin{equation}
\label{eq:ideal_gas}
\overline{P} =\mathcal{R}\, \overline{\rho}\, \overline{T},
\end{equation}
where $\mathcal{R}$ is the ideal gas constant, and where \(\overline{P}\), \( \overline{\rho} \), and \(\overline{T}\) denote the background pressure, density, and temperature, respectively.  We adopt the convention that an overbar indicates spherically-symmetric, time-independent background-state quantities.  The absence of an overbar in turn indicates a time-dependent and spatially varying perturbation about the background state.  As these perturbations relative to the background state are assumed to be small, we can linearize Equation \ref{eq:ideal_gas} to yield
\begin{equation}
\label{eq:linearized_eos}
\frac{\rho}{\overline{\rho}} = \frac{P}{\overline{P}} - \frac{T}{\overline{T}} = \frac{P}{\gamma \overline{P}} - \frac{S}{c_p},
\end{equation}
where $S$ is the specific entropy perturbation, $c_p$ is the specific heat at constant pressure and $\gamma$ is the adiabatic index. 

The anelastic continuity equation is given by
\begin{equation}
\label{eq:continuity}
     \bvec{\nabla}\cdot (\overline{\rho} \bvec{v}) = 0, 
\end{equation}
where $\bvec{v}$ is the velocity and where we adopt the convention that bold symbols denote vector quantities.  The evolution of $\bvec{v}$ is described by the momentum equation
\begin{equation}
\label{eq:momentum}
   \frac{\partial \bvec{v}}{\partial t} +\bvec{v}\cdot\bvec{\nabla}\bvec{v} + 2 \Omega_0 \hat{\bvec{z}} \times \bvec{v} = - \bvec{\nabla} \left(\frac{P}{\overline{\rho}}\right) + \frac{S }{c_p}g \bvec{\hat{r}} + \frac{1}{\overline{\rho}}\bvec{\nabla} \cdot \bvec{D}, 
\end{equation}
where $\Omega_0$ is the rotation rate of the star, $\hat{\bvec{z}}$ is the unit vector parallel to the rotation axis and $\bvec{g}$ is the gravitational acceleration.  The viscous stress tensor is indicated by $\bvec{D}$, and it is defined as
\begin{equation}
  D_{ij} = 2\, \overline{\rho}\, \nu \left( e_{ij} - \frac{1}{3} (\bvec{\nabla} \cdot \bvec{v}) \delta_{ij} \right),  
\end{equation}
where $e_{ij}$ is the strain rate tensor, $\nu$ is the kinematic viscosity and $\delta_{ij}$ is the Kronecker delta.  This system of equations is completed by a thermal energy equation describing the evolution of the specific entropy S,
\begin{equation}\label{eq:entropy}
    \overline{\rho} \overline{T}\left( \frac{\partial S}{\partial t} +\bvec{v}\cdot\bvec{\nabla}S \right) = \bvec{\nabla} \cdot (\overline{\rho} \overline{T} \kappa \bvec{\nabla} S) + \Phi,
\end{equation}
where $\kappa$ is the thermal diffusivity and the viscous heating term, $\Phi$, is defined as
\begin{equation}\label{eq:viscousheating}
\Phi = 2 \overline{\rho} \nu \left( e_{ij} e_{ij} - \frac{1}{3} (\bvec{\nabla} \cdot \bvec{v})^2 \right).
\end{equation}

\subsection{Numerical Algorithm}
For all simulations \postref{carried out} in this study, the Rayleigh convection code \postref{(v 1.2.0)}  was used to evolve Equations \ref{eq:ideal_gas}--\ref{eq:entropy} in 3-D spherical geometry \citep{Featherstone2016a,Matsui_etal_2016,Featherstone2024}.  Within \postref{the Rayleigh code}, the discretization and differentiation of system variables is accomplished using a spectral transform approach as described in \citet{Glatzmaier1984}.  In the radial direction, system variables are represented by a truncated expansion of Chebyshev polynomials $T_n(r)$ extending to maximum degree $n_\mathrm{max}$.  On spherical surfaces, all variables are expanded in a truncated series of spherical harmonic functions \postref{$Y_\ell^m$}, extended up to maximum degree $\ell_\mathrm{max}$.  All grids are dealiased, such that the number of collocation points in each dimension exceeds the number of spectral coefficients by a factor of 3/2.  Specifically, the number of radial and latitudinal collocation points, $N_r$ and $N_{\theta}$, are defined such that

\begin{equation}
    n_\mathrm{max} + 1 = \frac{2}{3} N_r \quad \text{and} \quad \ell_\mathrm{max} + 1 = \frac{2}{3} N_{\theta}.
\end{equation}

Time-integration is carried out using the semi‑implicit Crank–Nicolson scheme for linear operators and the explicit Adams-Bashforth method for nonlinear terms.  Finally, the divergence-free constraint (\ref{eq:continuity}) is enforced by decomposing the mass flux into poloidal and toroidal stream functions \postref{such} that:
\begin{equation}\label{streamfunction}
    \overline{\rho} \bvec{v} = \bvec{\nabla} \times \bvec{\nabla}\times  (W \bvec{\hat{r}}) + \bvec{\nabla} \times (Z \bvec{\hat{r}})
\end{equation}
where $\bvec{\hat{r}}$ is the radial unit vector, W and Z represent the poloidal and toroidal stream functions, respectively.  Additional details can be found in \citet{Featherstone2016a} and in \postref{the Rayleigh code documentation} (\url{https://rayleigh-documentation.readthedocs.io/en/latest/index.html}).

\subsection{Model Setup}
\label{sec:modelsetup}

\subsubsection{Polytropic Background State}
\postref{The simulations we carried out consist of 37 models.}  For each model, we adopt a polytropic thermal background state based on the prescription provided in \citet{Jones2011}.  As discussed in \citet{Featherstone2016a}, an appropriate choice of the polytropic parameters can yield a background state that closely resembles the standard solar model across much of the convection zone \citep[e.g.,][]{JCD1996}.  A summary of the polytropic parameters used to define the background state is provided in Table \ref{table:poly}.

\postref{Those} parameters are the same for all models presented here, with the exception of the outer radius.  For the bulk of our models, the outer radius is set to \postref{$r_\mathrm{o}=6.586\times10^{10}$} cm, and the thermal background closely resembles that of the solar convection zone between the same bounds in radius.  As we are motivated by dynamics of the solar near-surface shear layer, a relatively thin region of the upper convection zone, we also construct a set of models with half the shell thickness and a corresponding outer radius \postref{$r_\mathrm{o}=5.793\times10$} cm.

\tablepolytwo

\subsubsection{Boundary and Initial Conditions}
All models presented in this study were initialized with random thermal perturbations and zero fluid velocity.  They were then evolved for multiple viscous and thermal diffusion times following the onset of a statistically-steady state for the kinetic energy density and the energy flux balance (as defined in \S \ref{sec:global}). 

Conservation of mass and angular momentum were ensured by adopting stress-free and impenetrable boundary conditions at both the inner and outer boundaries such that
\begin{equation}
    v_r = \frac{\partial}{\partial r} \left( \frac{v_\theta}{r} \right) = \frac{\partial}{\partial r} \left( \frac{v_\phi}{r} \right) = 0
\end{equation}
\postref{at $r=r_i$ and $r=r_o$.} Our thermal boundary conditions differ between the \postref{outer and inner} boundaries.  At the \postref{outer} boundary, we enforce
\begin{equation}
S(r_o)=0.    
\end{equation}
This choice allows the conductive flux to vary spatially at the boundary, with the conductive flux $F_\mathrm{cond}$ defined as
\begin{equation}
F_\mathrm{cond}=-\overline{\rho}\overline{T}\kappa\frac{\partial S}{\partial r}.
\end{equation}
At the \postref{inner boundary}, we specify a fixed entropy gradient such that the incoming conductive flux varies in latitude.  The form we select for the entropy gradient is given by
\begin{equation}\label{lower-boundary}
\frac{\partial S}{\partial r} = - \frac{(1 + A Y_2^0)}{\rho T \kappa} \frac{L_\star}{4\pi r_i^2}
\end{equation}
\postref{at $r=r_i$. Here,} $L_\star$ is the luminosity imposed on the system, $Y_2^0$ is the $m=0$, $\ell=2$ spherical harmonic function, and $A$ is an amplitude factor that is varied across our suite of simulations.  The functional form of the lower boundary condition is illustrated in Figure \ref{fig:heatflux-bottom} for a range of $A$ values.  The sign of $A$ is chosen so that the imposed flux varies in latitude with the opposite sense of that which would be established when $A=0$.  Thus, for systems that evince an antisolar differential rotation and cool polar regions when $A=0$, we examine the impact of positive values for $A$.  Negative values of $A$ are explored for the handful of systems with solar-like differential rotation that we consider in this study.

\lowerbc

\subsubsection{Nondimensional Parameterization}\label{sec:nondim}
Throughout this paper, we describe some results in terms of the nondimensional parameters \postref{that characterize} a rotating, convective shell.  These nondimensional parameters express ratios of the relative timescales inherent to the system.  For instance, the viscous ($\tau_\nu$) and thermal diffusion ($\tau_\kappa$) timescales can be defined as 
\begin{equation}
\label{eq:tau_visc}
\tau_\nu \equiv \frac{L^2}{\nu}
\end{equation}
and
\begin{equation}
\label{eq:tau_kappa}
\tau_\kappa \equiv \frac{L^2}{\kappa}
\end{equation}
respectively, where $L$ is the depth of the fluid shell.  In addition, we have a rotational timescale given by
\begin{equation}
\label{eq:tau_omega}
\tau_\Omega \equiv \frac{1}{2\Omega_0}
\end{equation}
and a freefall timescale which we define as
\begin{equation}
\label{eq:tau_buoy}
\tau_{ff}\equiv\sqrt{\frac{L}{g'}}.
\end{equation}
Here, $g'$ is a characteristic value of the gravitational acceleration due to buoyancy, and it can be related to the imposed flux and the background state as described in \citet{Camisassa2022}.  These four timescales can in turn be combined to form the nondimensional control parameters that characterize rotating, convective systems such as those presented here.  Specifically, we have the Prandtl number, \postref{Pr,} given by
\begin{equation}
    \label{eq:Prandtl}
    \mathrm{Pr} = \frac{\nu}{\kappa} =\frac{\tau_\kappa}{\tau_\nu},
\end{equation}
which expresses the relative strength of viscous and thermal diffusivity.  For this study, we have chosen to fix the Prandtl number at unity.  We next have the Ekman number\postref{, Ek,} which expresses the relative importance of rotational and viscous effects, given by 
\begin{equation}
    \label{eq:Ek}
    \mathrm{Ek}=\frac{\nu}{2\Omega_0 L^2} = \frac{\tau_\Omega}{\tau_\nu}.
\end{equation}
Similarly, the competition between buoyancy and diffusion is expressed via the flux Rayleigh number, Ra$_\mathrm{F}$.  It can be defined as
\begin{equation}
    \label{eq:RaF}
 \mathrm{Ra}_\mathrm{F}=\frac{\widetilde{g} \widetilde{F} L^4}{c_p \widetilde{\rho} \widetilde{T} \nu \kappa^2} =\frac{\tau_\nu}{\tau_{ff}}\frac{\tau_\kappa}{\tau_{ff}},
\end{equation}
where the tildes indicate input parameters that have been volume averaged over the fluid shell \citep[c.f., ][]{Featherstone2016a,Camisassa2022}.
These input parameters can in turn be combined to yield the convective Rossby number, Ro$_\mathrm{c}$.  It describes the relative importance of the Coriolis and buoyancy forces; it is defined as
\begin{equation}
    \label{eq:Roc}
    {\rm Ro_c} \equiv \sqrt{\frac{\rm Ra_F\,Ek^2}{\rm Pr}} = \frac{\tau_{\Omega}}{\tau_{ff}}.
\end{equation}

It can also be useful to describe these systems in terms of two output parameters that characterize the resulting dynamics.  For each model in this study, we quote a Reynolds number, Re.  We choose a system-scale definition of Re such that
\begin{equation}
\label{eq:Re}
\mathrm{Re}=\frac{\rm \widetilde{U}\,L}{\nu},
\end{equation}
where $\widetilde{U}$ denotes an rms average over the fluid shell of the convective (i.e., non-asymmetric) flow amplitude.  We similarly quote the system-scale Rossby number, Ro, which is defined as
\begin{equation}
    \label{eq:Ro}
    {\rm Ro} = {\rm Re\, Ek} = \frac{\rm \tilde{U}}{\rm 2\Omega L}.
\end{equation}
Note that Ro is a measured output of the system and should not be confused with the \textit{convective Rossby number}, Ro$_\mathrm{c}$, which is an input control parameter.  A summary of input and output parameters for each model in this study is provided in Table \ref{tab:input-output}.

\section{Results}
\label{sec:results}

\subsection{Global Energetics}\label{sec:global}

\fluxbalanceradius

We first examine how the distribution and transport of energy responds to an imposed flux at the lower boundary.  We denote the kinetic energy density of the system by KE, which is defined as
\begin{equation}\label{eq:ke}
\mathrm{KE} = \frac{1}{2} \overline{\rho} \left( v_r ^2 + v_\theta^2 + v_\phi^2 \right).
\end{equation}
We can decompose the KE into contributions from the axisymmetric and non-axisymmetric components of the flow.  Doing so, we can define the convective kinetic energy (CKE) associated with non-axisymmetric motions as
\begin{equation}\label{eq:cke}
\mathrm{CKE} = \frac{1}{2} \overline{\rho} \left[ (v_r - \langle v_r \rangle)^2 + (v_\theta - \langle v_\theta \rangle)^2 + (v_\phi - \langle v_\phi \rangle)^2 \right].
\end{equation}
Here, the angled brackets denote an average over longitude.  The energy associated with axisymmetric meridional circulations (MCKE) in the $r-\theta$ direction is given by
\begin{equation}\label{eq:mcke}
\mathrm{MCKE} = \frac{1}{2} \overline{\rho} \left( \langle v_r \rangle^2 + \langle v_\theta \rangle^2 \right).
\end{equation}
Similarly, the energy associated with differential rotation (DRKE) can be defined as
\begin{equation}\label{eq:drke}
\mathrm{DRKE} = \frac{1}{2} \overline{\rho} \langle v_\phi \rangle^2.
\end{equation}

We find that it is primarily the axisymmetric motions that are impacted by an imposed flux variation at the lower boundary.  Figure \ref{fig:E_v_rsadius_shell_Avgs_} illustrates one representative example.   There we have plotted radial profiles of the different contributions to KE, averaged in time and over spherical surfaces, for a model with and without an imposed flux.  The CKE remains nearly unchanged by the imposed flux.  The energy in the mean flows, however, responds strongly to the imposed flux.  This is particularly true for the differential rotation, with DRKE decreasing by an order of magnitude between the two cases.  As is evident from Table \ref{tab:input-output}, we find this diminishment of DRKE in the presence of thermal forcing to be a consistent trend across our survey of models.  The one exception is model L5, in this initially solar-like case, we found that a forcing amplitude of $A=-1$ was sufficiently strong to not only diminish the differential rotation, but to reverse its sense entirely in the polar regions.  This reversal corresponds \postref{to} an enhancement in DRKE.

The relative insensitivity of the convective, non-axisymmetric motions to an imposed flux variation is also apparent when examining the flux of energy across the shell.  In a statistically steady state, the combined thermal and kinetic energy flux $\mbvec{F}$ satisfies

\begin{equation}\label{eq:total_flux}
    \bvec{\nabla}\cdot\mbvec{F}=\sum_i \bvec{\nabla}\cdot\postref{\mbvec{F}_i}=0,
\end{equation}
where the \postref{$\mbvec{F}_i$} indicate the contributions to the total flux budget by the enthalpy flux\postref{, $\mbvec{F}_{\rm e}$}, the conductive flux\postref{, $\mbvec{F}_{\rm c}$}, the kinetic energy flux\postref{, $\mbvec{F}_{\rm k}$,} and the viscous energy flux\postref{, $\mbvec{F}_\nu$}.  These contributions to the flux are defined respectively by
\begin{equation}\label{eq:enthalpy_flux}
      \mbvec{F}_{\rm e} = \bvec{v}\overline{\rho}\,\overline{T} c_p \left(\left(1-\frac{1}{\gamma}\right)\frac{P}{\overline{P}}+\frac{S}{c_p} \right)  ,
\end{equation}
\begin{equation}
         \mbvec{F}_{\rm c}(r) =- \kappa \overline{\rho}\,\overline{T}\bvec{\nabla}S,
\end{equation}
\begin{equation}
\mbvec{F}_{\rm k} =\frac{1}{2}\overline{\rho}\bvec{v} \left|\bvec{v} \right|^2
\end{equation}
and
\begin{equation}\label{eq:viscous_flux}
    \mbvec{F}_{\nu} =- \bvec{v}\cdot\bvec{D}.
\end{equation}

\fluxbalancetime

Figure \ref{fig:shell_Avgs_lowflow} illustrates one example of the flux balance achieved in a model with and without an imposed boundary flux variation, \postref{with $A=0.5$ and $A=0$ respectively}. There, we have plotted the radial component of each contribution to the total flux\postref{, $\mathcal{F}_{i,r}$,} averaged over time and spherical surfaces at each radius.  In addition, we have converted these fluxes to a luminosity by multiplying by $4\pi r^2$.   We see that, outside of the thermal boundary layers, the dominant balance is struck between enthalpy and kinetic-energy flux for both cases.  In addition, we do not see any substantial difference in the shape of the different curves when comparing the unforced case (panel $a$) to the case with a variable, lower-boundary flux (panel $b$).  

This result is consistent with Figure \ref{fig:E_v_rsadius_shell_Avgs_}, which suggests that the mean flows are most substantially impacted by an imposed lower-boundary flux.  Of those, only the meridional flow can transport energy in radius but, because that flow is largely north-south antisymmetric, its contribution to the averages depicted in Figure \ref{fig:shell_Avgs_lowflow} is minimal.

\subsection{Latitudinal Response of the Mean Flow and Thermal Profiles}\label{sec:meanflows}
\entropyDR

We next examine how the differential rotation and thermal profiles respond to an imposed heat flux. \postref{In Figure \ref{fig:Entropy_DR}, we illustrate how the radial and latitudinal variation of the axisymmetric entropy perturbation, $\langle S\rangle$, and differential rotation, $\langle \Omega - \Omega_0 \rangle$, respond to variation of the forcing amplitude $A$.}  In the absence of any imposed flux, the \hrc ~model develops cool polar regions at the surface, a warm equatorial region, and an antisolar differential rotation.  The opposite is true for the \lrc ~model.  This behavior is consistent with that expected for systems with Ro$_\mathrm{c} > 1$ and Ro$_\mathrm{c} < 1$ respectively \citep[e.g.,][]{Gastine2014,Camisassa2022}.

 As the sign of $A$ has been chosen to oppose the latitudinal entropy variations that each system would naturally establish, we expect to observe changes in the mean thermal structure and, possibly, the mean flows as $A$ is increased.  Such a trend is evident in Figure \ref{fig:Entropy_DR}.  As the forcing amplitude is increased, the latitudinal entropy variation in the \hrc ~model diminishes until, at sufficiently high $A$, a thermal profile of the opposite sense in latitude is established.  This behavior is similar for the \lrc ~model as illustrated in the panels at right. 

As the latitudinal thermal perturbations change, we observe a corresponding change in the differential rotation.  In both the high- and \lrc ~systems,  the differential rotation weakens as the amplitude of $A$ is increased. In the \hrc ~model, we can observe spinning down of the polar regions, suggesting that the system is transitioning toward a solar-like state of differential rotation.  In the case of the \lrc ~model, an amplitude of $A=-1$ leads to a complete reversal of the differential rotation in the polar regions.

In summary, a non-uniform flux imposed at the base of the convection zone can drive the axisymmetric response of the system toward a state that has the opposite sense of that achieved with no forcing.   As we discuss in the next section, this behavior seems to result from the modification of the thermal wind balance established in the presence of forcing.

\subsection{Implications for Thermal Wind Balance}\label{sec:TWB}
When inertial, viscous and magnetic effects can be neglected, the dominant force balance in a rotating convection zone is instead struck between the buoyancy, pressure and Coriolis forces.  This balance, known as thermal wind balance, appears to arise in the bulk of the solar convection zone where it has been used to explain the non-cylindrical contours of differential rotation \citep[e.g., ][]{Rempel2005,Miesch2006,Matilsky2020}.   The assumption of thermal wind balance is also a central to the model of \citet{Jha2021} that reproduces the structure of \postref{the} near-surface shear layer.   

\thermalwindboth

We examine thermal wind balance in our models by considering the $\phi$-component of the curl of the momentum equation. When inertial and viscous forces are neglected, we have that  
\begin{equation}\label{eq:twb}
   \frac{g}{r c_{p}} \frac{\partial \langle S \rangle}{\partial \theta} = 2 \Omega_{0} r \sin\theta \frac{\partial \langle \Omega \rangle}{\partial z}.
\end{equation}
The left-hand side of this equation represents baroclinic forcing due to latitudinal entropy gradients.  \postref{That forcing is balanced on the right-hand side by the Coriolis force since the gradient of reduced pressure vanished when the curl was taken}. 

One question this study \postref{seeks to answer} is the extent to which latitudinal heat flux variations imposed at the base of the convection zone can influence thermal wind balance in the high- and \lrc regimes.  The equilibrated regime for several high- and \lrc ~models is illustrated in Figure \ref{fig:thermal_wind_high_Ro}.  When thermal wind balance holds, the Coriolis contribution (upper panel) and the baroclinic contribution (lower panel) closely align.

In the absence of an imposed flux, our results show that thermal wind balance is well-satisfied in the low-Ro$\mathrm{c}$ regime, while a small departures occur in the high-Ro$\mathrm{c}$ regime.  This is evident from Figure \ref{fig:thermal_wind_high_Ro}, which depicts the close correspondence between the upper and lower panels, \postref{particularly within} the low-Ro$\mathrm{c}$ regime.  By contrast, in the high-Ro$_c$ regime, there is good agreement in the equatorial regions, but there are notable  departures in the high-latitude regions. 

In both the \hrc ~and \lrc ~systems,  the presence of an imposed flux variation does not lead to a loss of thermal wind balance.  It does lead, however, to a change in the form of the balance achieved.  For both systems, the imposed flux significantly weakens both the baroclinic and Coriolis terms in the equatorial regions.  In the polar regions, however, their strength increases, and both the baroclinic and Coriolis terms largely maintain \postref{the radial and latitudinal structure established} in the absence of any forcing.    

\entropydrmc

\subsection{Effect of Increasing Rossby Number at Fixed Forcing}
\label{sec:thickshell}

\thermalwindvaryro

So far, we have examined how a variation in the amplitude of an imposed flux can affect the flows and balances achieved.   We now examine how variations in Ro$_\mathrm{c}$, at fixed value of $A$, influence the mean flows and heat transport.  Here, we fix the value of $A=0.3$ and, motivated by the NSSL, focus on systems operating in the \hrc ~regime.  We have also run a series of models with half the shell thickness of those examined so far, similarly fixing $A=0.3$ and varying Ro$_\mathrm{c}$.  With the exception of the emergent flux, which we discuss at the end of this section, we observe qualitatively similar behavior between the thin-shell models and the thick-shell models.

We observe that as Ro$_\mathrm{c}$  increases, and the system transitions to an increasingly buoyancy-dominated regime, it ultimately loses any large-scale organization in its axisymmetric flows and fields, which instead become dominated by small-scale structures.  Figure \ref{fig:entropydrmc} illustrates this trend.  There, we have plotted longitudinal and time averages of differential rotation, meridional circulation and specific entropy perturbation for three representative values of Ro$_\mathrm{c}$.  We find that, for $A=0.3$, a transition point is crossed by Ro$_\mathrm{c}\approx7$.  For lower values of Ro$_\mathrm{c}$, we find that the resulting profiles are consistent with the buoyancy-dominated regime.  As Ro$_\mathrm{c}$ is increased toward the transitional value of 7, we observe only minimal changes in the mean-flow profiles. \postref{This can be seen, for instance, by comparing the $A=0.3$, Ro$_\mathrm{c}=1.75$ model from Figure \ref{fig:Entropy_DR} with the Ro$_\mathrm{c}=3.5$ model of Figure \ref{fig:entropydrmc}}.  The most notable of these is a tendency for differential rotation profiles weaken and to deviate from rotation on cylinders. 

Around Ro$_\mathrm{c}\approx 7$, we observe that the differential rotation loses its anti-solar structure entirely.  Instead, the inner portion of the convective shell rotates more rapidly than the outer portion, and the differential rotation shows only weak variation in latitude (Figure \ref{fig:entropydrmc}$b$).  The change in differential rotation is accompanied by a reversal in the sense of the meridional circulation, which becomes equatorward at the surface.  

\postref{The reversal of meridional circulation likely results} from the enhancement of the latitudinal thermal perturbations evident in the lower panel of Figure \ref{fig:entropydrmc}, combined with the loss of rotational constraint.   The imposed warm poles \postref{force} upwelling circulation in the polar regions, and the cool equator similarly drives downwelling flows at low latitudes.  This effect can be offset through the conservation of angular momentum, which tends to drive circulation in the opposite direction \postref{through} gyroscopic pumping.  \postref{That} processes, however, becomes weaker as the relative importance of the Coriolis force decreases \postref{as discussed in} \citet{Miesch2011, Featherstone2015}.  As the value of Ro$_\mathrm{c}$ is increased even more, the effects of the thermal forcing also begin to diminish, and we observe a general loss of symmetry about the equator for both the mean flows and thermal structure.  

The general trend observed in the mean flows is similarly reflected in variations that occur in the thermal wind balance as Ro$_\mathrm{c}$ is increased.  These changes are illustrated in Figure \ref{fig:thermal_wind_vary_Ro}.  For $A=0.3$, the structure of the baroclinic and Coriolis terms is reminiscent of the \hrc, unforced \postref{model} shown in Figure \ref{fig:thermal_wind_high_Ro}.  As Ro$_\mathrm{c}$ is increased, the structure of those terms resembles the strongly forced systems of Figure \ref{fig:thermal_wind_high_Ro}.  Beyond Ro$_\mathrm{c}=7$, that structuring, which results from the imposed latitudinal flux variation, becomes increasingly weaker and more disordered.  

\postref{This suggests} that sufficiently high-Ro flows can efficiently mix lateral variations in heat that are set from below.  We can observe this effect more directly by considering the outward flux of thermal energy at the upper boundary.  The resulting outward heat flux is also found to change significantly as Ro$_\mathrm{c}$ is increased, with the change being most notable for the thick-shell series of models.  

In Figure \ref{fig:heat_flux_top}$a$, we plot the time- and longitudinally-averaged upper-boundary heat flux for our thick-shell, \hrc ~models with forcing amplitude $A=0.3$.   As we have chosen to enforce $S=0$ at the upper boundary, its gradient (and hence the conductive flux) is free to vary at that boundary.  At lower values of Ro$_\mathrm{c}$ the profile of emergent flux reflects that of the imposed flux, with the high latitudes exhibiting enhanced flux with respect to the equator.  Beyond the transitional value of Ro$_\mathrm{c}=7$, however, the resulting heat flux becomes nearly uniform in latitude.

 This trend also holds true for those models that were run using a convective shell with half the thickness (Figure \ref{fig:heat_flux_top}$b$). The primary difference we observe in those models is that the surface flux does not become as uniform at low latitudes as in the thick-shell cases. These differences are marginal, however, and may be related to the fact that the conductive thermal boundary layer occupies a greater fraction of the convective shell \postref{in these models}.  To test this, we ran a second series of thin shell models at \postref{lower Ra$_\mathrm{F}$ in order to produce a thicker boundary layer.  However, we observed only small differences in the surface flux distribution when compared to the series of models run at higher Ra$_\mathrm{F}$ ( Figure \ref{fig:heat_flux_top}$c$).}  We suggest it is possible that either we have not considered a wide enough range of Ra$_\mathrm{F}$ to observe a significant change, or the differences between the thin and thick shell models stem from differences in the connectivity of large-scale flows that arise due to changes in the shell aspect ratio.

\resultheatflux

\inputoutput

\section{Summary and Discussion}
\label{sec:summary}



In this study, we have explored how modifications to thermal wind balance can impact the establishment of mean flows and the efficiency of heat transport in a rotating convective system.  This was accomplished by imposing an axisymmetric, but latitudinally-varying heat flux at the lower boundary of a series of solar-like convection zone models operating in parameter regimes that are both rotationally-constrained and rotationally-unconstrained.  In some respects, these results build on the work of \citet{Miesch2006} and \citet{Matilsky2020} that focused on differential rotation in the bulk of the solar convection zone.  Those studies explored how boundary-induced modifications to thermal wind balance, which generally strengthened that already present in the system,  can drive small changes in the isosurfaces of differential rotation established in low-Ro systems.

Our study explores how boundary-induced modifications \postref{that oppose} the naturally-established thermal wind balance impact the properties of both low-Ro and high-Ro systems. This approach was motivated by the fact that while the differential rotation of the bulk convection zone suggests that it operates in a low-Ro regime, observations of the NSSL clearly indicate the flows there are high-Ro in nature.  And yet, the NSSL does not evince an antisolar differential rotation or polar regions that are measurably cooler than the equator.  Moreover, analytical models that successfully reproduce its structure suggest that it operates in thermal wind balance, a state typically associated with low-Ro systems \citep[][]{Choudhuri2021, Jha2021}.  By imposing a latitudinally-varying heat flux that opposes that which is established in the absence of any forcing, we have been able to explore how a convective layer might respond to a thermal wind balance that is established below its base.  

We have found that the changes induced in the mean flows can be significant, provided that the thermal forcing is sufficiently strong.  This is ready illustrated in Figure \ref{fig:Entropy_DR} where we see that as the amplitude $A$ of the imposed flux variation is increased, the differential rotation weakens.  \postref{ In fact, for sufficiently large values of $A$, the differential rotation develops a latitudinal variation that is reversed with respect to that achieved in the absence of any forcing.}  

This change in the mean flows corresponds with a change in the thermal wind balance achieved (e.g., Figure \ref{fig:thermal_wind_high_Ro}).  In both the \lrc ~and \hrc ~regimes, the baroclinic and Coriolis contributions to the balance are weakened at the equator and considerably strengthened at the poles in response to an imposed flux variation.  In the absence of any forcing, we found that both \lrc ~\textit{and} \hrc ~systems exhibited a high degree of thermal-wind balance.  These results may shed some light on why the differential rotation of the NSSL is not observed to be antisolar.  They also support the plausibility of the key assumption of the model of \citet{Choudhuri2021} and \citet{Jha2021}, namely that the high-Ro flows of the NSSL are in thermal wind balance.

The effects of the imposed flux ultimately depend not just on its amplitude, but also on how strongly the convection is influenced by the Coriolis force.  At sufficiently \hrc ~we find that even in the presence of an imposed flux, a rotating system cannot maintain thermal wind balance (Figure \ref{fig:thermal_wind_vary_Ro}).  When this happens, lateral variations in entropy that would be established by the lower boundary forcing are well-mixed, so that the outer boundary of the model exhibits no large-scale variations in heat flux with latitude \postref{(Figure \ref{fig:heat_flux_top})}.  \postref{We suggest it is possible that such an effect may contribute to the very weak latitudinal temperature variations observed in the solar photosphere} \citep{Kuhn1998, Kuhn1998b, Rast2008}.


It is worth noting that the principle utility of the models presented here, namely their simplistic design, also represents a fundamental limitation.  The models in this study capture only the thermal imprint of deep convection for instance, and not that of its differential rotation or meridional circulation.   They are also nonmagnetic, whereas the numerical simulations of \citet{Hotta2025}, which self-consistently reproduce a near-surface layer of shear, indicate that the presence of Lorentz forces can produce significant departures from thermal wind balance in the upper convection zone.  We plan to explore the impacts of both magnetism and an imposed flow in future work.  

\postref{Finally, we remark that while the Sun is comprised of turbulent, high-temperature plasma, this is not true of other objects in our solar system.  And yet,} many moons are thought to possess subcrustal liquid oceans that convect due to secular cooling and tidal heating of the core/mantle at their base \citep[e.g.,][]{Nimmo2016,Soderlund2024}.  Tidal heating in particular can lead to spatial inhomogeneities in the convective heat flux reminiscent of those considered here and which may ultimately impact the surface ice shell thickness, as may be the case for Europa \citep[e.g.,][]{Lemasquerier2023}.  And so, while motivated by the Sun, these results might also \postref{relate to convective systems in other solar-system bodies}.



This project was primarily supported by NASA grant 80NSSC22M0162 (COFFIES DRIVE Center) and NSF grant 2405049 (SHINE).  N. Featherstone received additional support through NASA grants  80NSSC24K0125 (HTMS) and 80NSSC24K0271 (HSR).  Computational resources were provided by NASA’s High-End Computing (HEC) program through the Pleiades supercomputer. The Rayleigh code has been developed with support by the National Science Foundation through the Computational Infrastructure for Geodynamics under grants NSF-0949446 and NSF-1550901.  We are grateful to Kyle Augustson and Brad Hindman for several insightful discussions related to this work.

\bibliography{paper}{}
\bibliographystyle{aasjournalv7}
\end{document}